\documentclass[fp]{jpsj3}
\usepackage{txfonts}
\usepackage{setspace}

\title{
Collective Adoption of Max-Min Strategy in an Information Cascade Voting
Experiment
}

\author{Shintaro Mori$^1$\thanks{E-mail: mori@sci.kitasato-u.ac.jp},
Masato Hisakado$^2$, and Taiki Takahashi$^{3}{}^{,}{}^{4}$}
\inst{$^1$Department of Physics, Kitasato University \\ 
1-15-1 Kitasato, Sagamihara, Kanagawa 252-0373, Japan \\
$^2$Standard and Poor's \\
 1-6-5 Marunouchi, Chiyoda-ku, Tokyo 100-0005, Japan \\
$^3$
Department of Behavioral Science, Faculty of Letters
\\
and
\\
Center for Experimental Research in Social Sciences, Hokkaido University
$^4$
\\
Kita 10, Nishi 7, Kita-ku, Sapporo, Hokkaido 060-0810, Japan} 

\abst{
\setstretch{1.0}
 We consider a situation where one has to choose an option with
 multiplier $m$. The multiplier is inversely proportional to the number of
 people who have chosen the option and is proportional to the return if
 it is correct.  
 If one does not know the correct option, we call him a herder, and then 
 there is a zero-sum game between the herder and 
 other people who have set  the multiplier.  
 The max-min
 strategy where one divides 
 one's choice inversely proportional to $m$ is optimal   
 from the viewpoint of the maximization of expected return.
 We call the optimal herder an analog herder.
 The system  of analog herders takes the probability of 
 correct choice to one for any value of the ratio of
 herders, $p<1$, in the thermodynamic limit if the
 accuracy of the choice of informed person $q$ is one. 
 We study  how herders choose  
 by a voting experiment 
 in which 50 to 60 subjects sequentially answer a two-choice quiz.
 We show that the probability of selecting a choice by the herders
 is inversely proportional to $m$ for 
 $4/3 \le m \le 4$ 
 and they  collectively  adopt 
 the max-min strategy in that range.
}

\kword{herd, information cascade, zero-sum game, experiment, max-min
strategy, econophysics, socio-physics}

\begin{document}
\maketitle

\setstretch{1.0}

\section{\label{sec:intro}Introduction}
 Even if each person has limited information, aggregated information 
becomes very accurate \cite{Smi:1996}. This is the wisdom of crowd
effect, and is
supported by many examples from political elections, sports predictions,
quiz shows, and prediction markets \cite{Sur:2004,Pag:2008,Mil:2011}. In
contrast, in order to give
accurate results, three conditions need to be satisfied: 
diversity, independence, and decentralization. If these conditions are
not satisfied, aggregated information becomes unreliable or worse 
\cite{Sur:2004,Lor:2011}. 
However, in an ever-more connected world, it becomes more and more 
difficult to retain independence. Furthermore, if the 
actions or choices of others are visible, neglecting them is not 
 realistic in light of the merit of social
 learning \cite{Ren:2010,Ren:2011}. In this case, 
information cascade  may emerge  and  
information aggregation ceases 
\cite{Bik:1992,And:1997,Kub:2004,Goe:2007,Lee:1993,Dev:1996,Wat:2002}.  

More concretely, we consider a situation where people sequentially answer a
two-choice question with choices A and B.
  The payoff for the correct choice 
  is constant .
 Before this question is asked, 
many other people have already answered and their choices are made known
 as $C_{A}$ people choosing A and $C_{B}$ people choosing B, which 
 is called social information. 
If the person answering knows 
the correct choice, he should choose it. His choice is not affected by 
 social information.
 We then call him an independent voter.
  However, if he does not know the correct choice, 
 he will be affected by social information \cite{Lat:1981}.
He tends to go with the majority and we then call him a herder. By  
herding, the wisdom of crowds is on the edge. If a herder is 
isolated from others, his choice becomes A and B,
 and should be  canceled. As a result, 
 the  choice by an independent voter remains. 
 The majority choice always
 converges to the 
 correct one in the limit of a large 
 number of people. This is known as Condorcet's jury theorem \cite{Smi:1996}.
However, if others' choices are given as social
 information, 
the cancellation mechanism does not work. 
The herder copies the majority and ignores the correct information 
given by the independent voter. If the proportion of herders $p$ exceeds 
some threshold value $p_{c}$, there occurs a phase transition from the 
one-peak phase where the majority choice always converges to the correct one
to the two-peak phase where 
the majority choice converges to the wrong one with a finite and 
positive probability \cite{Mor:2012}.
We call this phase transition the information cascade 
transition \cite{His:2011,Mor:2012,His:2012}.
This is the risk of imitation in the wisdom of crowd. How can we  
 avoid this risk? There exists a hint in race-track betting markets \cite{Hau:2008,Ali:1977} and 
prediction markets \cite{Wol:2004,Man:2006}. In order to aggregate 
information scattered among people, the market mechanism can 
be very effective \cite{Sur:2004,Pag:2008,Mil:2011}.

 We consider a situation in which each choice $\alpha\in \{A,B\}$ 
 has a multiplier $M_{\alpha}$ that is inversely proportional to 
 the number of subjects $C_{\alpha}$ who chose it.
  The payoff for the correct choice 
  is proportional to the multiplier.
  If the multiplier of a choice is large, the number of people who 
  chose it is small.
  If the return is constant,  herder usually
   avoid the choice. 
  However, now,  the return on the correct choice is proportional to the
 multiplier, and hence 
  we cannot say that herder does not choose it. Copying the majority gives him a
  small return, even if it is a correct choice. The multiplier plays the 
  role of a ``tax'' on herding (free riding) and copying the minority can be an
  attractive choice. 
  The situation is a zero-sum game between the herder answering and all the
  previous subjects who have set the multipliers. In zero-sum games, 
  the max-min strategy maximizes the expected
  return and is optimal \cite{Neu:1944}. In the above two-choice quiz, 
  the max-min strategy is the one
  where a herder chooses $\alpha$
  with a probability proportional to $C_{\alpha}$ and cancels the risk
  in expected return  by the multipliers.
  We call the herder who adopts the optimal max-min strategy  
  an analog herder \cite{Mor:2010,His:2010}.
  If a herder behaves as an analog herder,
  the convergence to equilibrium state becomes 
  slow as $p$ increases and 
  there occurs a phase transition in
  the convergence speed as $p$ exceeds half \cite{His:2010}.
  However,
  the information cascade phase transition does not occur for $p<1$.
  A majority of people always choose the correct choice in the limit 
   of a large number of
  people (thermodynamic limit) and the system is in the one-peak 
  phase for any value of $p$ if the accuracy of 
  the information of the independent
  voter $q$ is $q>1/2$ \cite{His:2010}. 
  Furthermore, the analog herder's choice  does not affect 
  the limit value of the percentage of correct
  answer and it converges
  to $q$.   
  As for the two-choice quiz, the
  independent voter knows the correct choice and $q=1$ holds.
  In this case, the system of analog herders maximizes 
  the probability of correct choice for $p<1$
  in the thermodynamic limit. 
  Even in limit $p\to 1$, the system can take the probability
  to one.

 In this paper, we have adopted an experimental approach to study
  whether herders adopt the max-min strategy and 
  behave as analog herders if the choices have multipliers. 
 We have also studied a herder's probability of correct choice.
 The organization of the paper is as follows.
 We explain the experiment and derive the optimal max-min strategy in section
 \ref{sec:exp}.
 The subjects answer a two-choice quiz in three cases $r\in \{O,C,M\}$.
 In case $O$, the subjects answer without social information.
 In cases $C$ and $M$, they receive social information based on 
 previous subjects' choices.
 Social information is given as summary 
 statistics $\{C_{A},C_{B}\}$ in case C
  and as multipliers $\{M_{A},M_{B}\}$ in case M.
 Sections \ref{sec:analysis} and \ref{sec:analysis2} 
 are devoted to the analysis of the
  experimental data. In section \ref{sec:analysis}, we summarize
  data about  the macroscopic aspects of the system. 
 In section \ref{sec:analysis2}, we derive a microscopic rule regarding how 
 herders copy others in each case $r \in \{C,M\}$.
 In section \ref{sec:model}, we introduce a stochastic model that
 simulates the system. We study the transition ratio $p_{c}(r)$ 
  for cases $r\in \{C,M\}$.
  We estimate the probability of correct choice  by the 
  herders in the experiment and compare it with 
  that of the optimal analog herders system.
  Section \ref{sec:conclusions} is devoted to the summary and discussions. 
  In the appendices, we give some supplementary information about the
  experiment and 
  a simulation study of the convergence exponent. 
  We also prove that only the system of  analog herders can
  take the
  probability of 
  correct choice to one in limit $p\to 1$.

\section{\label{sec:exp}Experimental setup  and optimal strategy in case $M$}

\subsection{Experimental setup}
 The experiment reported here was conducted at the Group 
Experiment Laboratory of the 
 Center for Experimental Research in Social Sciences at Hokkaido
 University.
 We have conducted two experiments. We call them EXP-I and EXP-II.
 In EXP-I (II), we recruited  120 (104) students 
 from the university. We divided them into two groups, Group A and Group B,
 and prepared two sequences of subjects of average length 60 (52).
 The main motive to divide the subjects into two groups is to ensure many 
 choice sequences in order to estimate the average value of 
 macroscopic quantities.
 In addition, we can check the estimation
  of herders' ratio $p$ by comparing the values from the two groups 
  for the same question \cite{Mor:2012}.  

 The subjects sequentially answered a two-choice quiz of 120 questions.
 Some subjects could not answer all the questions 
 within the alloted time, and so   
 the number $T$ of subjects who answered a particular question varied.
 We label the questions by $i\in \{1,2,\cdots,120\}$ and denote the
 length of the sequence of the subjects for question $i$ by $T_{i}$. 
 In EXP-I, the subject answers in three cases $r \in \{O,C,M\}$ 
 in this order. We denote the answer 
 to question $i$ in case $r$ after $t-1$ subjects' answers by
 $X(i,t|r)$, which takes the value 1 (0) if the choice 
 is true (false). The order $t$ of the subject
 in the choice sequence $\{X(i,t|r)\}$ 
 plays the role of time.
 $\{C_{0}(i,t|r),C_{1}(i,t|r)\}$ are the number of
 subjects who 
 choose true and false for question $i$ among 
 the prior $t$ subjects and are given as
\begin{eqnarray}
C_{1}(i,t|r)&=&\sum_{t'=1}^{t}X(i,t'|r), \nonumber \\
C_{0}(i,t|r)&=&t-C_{1}(i,t|r) \nonumber . 
\end{eqnarray}
 In case $O$, the subject answered without 
 any social information.  
 Then, he answered in case $C$. When $t-1$ subjects have already answered 
 question $i$ before him  in his group, he received  
 summary statistics $\{C_{0}(i,t-1|C),C_{1}(i,t-1|C)\}$ from all of them. 
 For the correct choice in cases $O$ and $C$, the subject gets two 
 points. 
 Finally, in case $M$, when $t-1$ subjects have already answered 
 question $i$ before him in his group, 
 the subject receives multipliers
 $\{M_{0}(i,t-1),M_{1}(i,t-1)\}$ from all previous $t-1$ subjects. 
 For the correct choice, the subject gets the points which is given by the 
 multiplier. The multiplier $M_{\alpha}$ for $\alpha \in \{0,1\}$ was 
 calculated based on the summary 
 statistics in case $M$ as 
\begin{eqnarray}
M_{\alpha}(i,t-1)
&=&\frac{C_{0}(i,t-1|M)+C_{1}(i,t-1|M)+1}{C_{\alpha}(i,t-1|M)+1}
 \nonumber \\
&=&
\frac{t}{C_{\alpha}(i,t-1|M)+1} \nonumber .
\end{eqnarray}
 The multiplier is given by dividing total points $C_{0}+C_{1}+1=t$
 for all subjects with choice value 
 among $C_{\alpha}+1$ subjects who have chosen $\alpha$. 
 This is similar to the payoff odds of the 
 parimutuel system in gambling. 
 
 In EXP-II, in addition to the three cases $r\in \{O,C,M\}$,
 the subjects answered in at most four cases $r\in \{1,5,11,21\}$ 
 between cases $O$ and $C$. 
 In cases $r\in \{1,5,11,21\}$, the subject received summary
 statistics $\{C_{0}(i,t-1|r),C_{1}(i,t-1|r)\}$ from previous $r$
 subjects. $C_{0}(i,t-1|r)+C_{1}(i,t-1|r)=r$ holds and as $r$ increases,
 the amount of social information increases.
 In EXP-I, the amount of social information 
 increases rapidly from $r=0$ in case $O$ to $r=t-1$ in case $C$.
 In EXP-II, $r$ gradually increases. 
 The payoff for the correct choice is 1 in cases 
 $r\in \{O,1,5,11,21,C\}$ and the multiplier in case $M$.
 Detailed information about EXP-II has been presented in our
 previous work \cite{Mor:2012}, where we have studied the experimental data for
 cases $r\in \{O,1,5,11,21,C\}$. In this paper, we  concentrate on
 case $M$ and take case $C$ as the control case. 
 
 We repeated the same experiment for both Groups A and B. 
 We obtained $120\times 2$ sequences
 $\{X(i,t|r)\}$ for each $r \in \{O,C,M\}$. We label the sequence 
 in Group B by $i+120$, so that  
 $i \in \{1,2,\cdots,240\}$. The experimental design is summarized 
 in Table \ref{tab:design}. 

\begin{table}[htbp]
\caption{\label{tab:design}
\setstretch{1.0}
Experimental design. $T$ means the number of subjects and $\{r\}$ means the 
cases where the subjects answered the quiz. 
$I$ means the number of questions.
 The length $T_{i}$ of sequence $\{X(i,t|r)\}$ for question $i$
 is almost the same as $T$ in EXP-I. In EXP-II, it depends on $i$ and 
the average value is $50.8$.
}
\begin{tabular}{lcccc}
\hline
Experiment & Group & $T$ & Cases $\{r\}$ & $I$ \\ 
\hline
EXP-I  & A & 57 & $\{O,C,M\}$ & 120 \\
EXP-I  & B & 63 & $\{O,C,M\}$ & 120 \\
EXP-II & A & 52 & $\{O,1,5,11,21,C,M\}$ & 120  \\
EXP-II & B & 52 & $\{O,1,5,11,21,C,M\}$ & 120  \\
\hline
\end{tabular}
\end{table}

\subsection{\label{Max}Max-Min Strategy in case $M$}
 We derive the optimal strategy 
for herders in case $M$.  
A subject can choose $\alpha \in \{A,B\}$.
We suppose that he votes one unit for a choice and call him a voter.
Here, we consider  the case where one vote can be divided by the voter.
If a voter  believes $A$ is correct,  he  votes one unit for $A$.
If the voter does not know the answer at all,  he votes  0.5 unit for
$A$  and 0.5 unit for $B$.
We assume that a voter thinks  the probability that  $A$ is correct 
is  $\beta$, and  the  probability that $B$ is correct is  $1- \beta$.
The voter  divides one unit vote  into  $x$ for $A$ and  $1-x$ for $B$ 
by his decision making.
Expected return $R$ is 
\begin{eqnarray}
R&=& \beta \cdot M_{A} \cdot x+ (1-\beta) \cdot M_{B} \cdot (1-x) 
\nonumber  \\
&=&\beta(M_{A} x-M_{B}(1-x)) +M_B(1-x).
\label{b1}
\end{eqnarray}

We assume that herders   do not have information about the  correct 
answers without multipliers $\{M_{A},M_{B}\}$.
Hence, we assume  that a herder cannot estimate the probabilities  of correct 
answer  $\beta$ as Knightian uncertainty, because he has no
knowledge  to answer the question \cite{Kni:1921}.
The situation is a zero-sum game between the herder and other
previous voters as the multipliers are 
set such that all votes are divided by the voters who have chosen 
the correct option.
The max-min strategy has been proved to be optimal in 
game theory \cite{Neu:1944}.
The voter minimizes the expected loss due to the 
 uncertainty in the choice.
 In order to minimize the expected loss from the uncertainty,   
it should be chosen such that $M_{A}\cdot x  = M_{B}\cdot (1-x) $ 
holds, from (\ref{b1}).
This position has no sensibility for $\beta$.

We can calculate  $x$  from (\ref{b1}),
\begin{equation}
x=\frac{M_{B}}{M_{B}+M_{A}}.
\end{equation}
As multiplier $M_{\alpha}$ is calculated as
\[
M_{\alpha}=\frac{t+1}{C_{\alpha}+1},
\]
ratio $x$ for $A$ is then 
\begin{equation}
x=\frac{C_A+1}{t+2} \sim \frac{C_A}{t} \hspace*{0.3cm}
\mbox{for}\hspace*{0.3cm}t>>1.
\end{equation}
$x$ becomes proportional to $C_A$ and 
 it is the voting strategy of analog herders \cite{His:2010}. 

The discussion shows that the strategy of analog herders is
optimal for a herder as it  maximizes his expected return. 
 In our experiment, the voter cannot divide one's vote (choice).
 Hence, the averaged behavior of herders becomes akin to that of 
the analog herders,  when herders adopt the optimal  strategy.

We make one comment about the optimal strategy for the independent
voter. When $\beta=1$, the voter believes his information and chooses 
what he believes to be true.
When $\beta<1$, it is not optimal to do so in general.
The expected return $R$ in (\ref{b1}) is
\begin{equation}
R= (\beta \cdot M_{A}-(1-\beta)\cdot M_{B})x+M_{B}(1-\beta).
\label{B1}
\end{equation}
By maximizing $R$, we obtain $x$ as
\[ 
x=\theta(\beta\cdot M_{A}-(1-\beta)\cdot M_{B}).
\]
Here, $\theta$ is a Heaviside (step) function.
If $\beta\cdot M_{A}>(1-\beta)\cdot M_{B}$, he chooses $A$ and vice versa.
He behaves as an ``arbitrager'' for $\beta<1$.
It is the risk-neutral strategy that has been discussed 
in the context of racetrack betting markets and prediction markets
\cite{Ali:1977,Man:2006}.

\section{\label{sec:analysis}Data analysis : Macroscopic Aspects}
 We obtained 240 sequences
 $\{X(i,t|r)\},t\in\{1,2,\cdots,T_{i}\}$ 
 for question $i \in \{1,\cdots,240\}$ and 
 cases $r\in\{O,C,M\}$ in each experiment. 
  Data $\{X(i,t|r)\}$ for both experiments is downloadable at 
 http://arxiv.org/abs/1211.3193.
 The percentage of correct answers of sequence $\{X(i,t|r)\}$ for 
 question $i$ in case $r$ 
 is defined as  $Z(i|r)=\sum_{s=1}^{T_{i}}X(i,s|r)/T_{i}$. 
 In the analysis, the subjects are classified into two categories--independent voters and herders--for each question. 
 We assume that the probability $q$ of a  
 correct choice for independent voters and 
 herders is 100\% and 50\%, respectively \cite{Mor:2012}.
 For a group with $p(i)$ herders and $1-p(i)$ independent voters, the 
 expectation value of $Z(i|O)$ is $1-p(i)/2$. 
 The maximal likelihood estimate of $p(i)$ is 
 given as $p(i)=2(1-Z(i|O))$. 
\subsection{Distribution of $Z(i|r)$}
\begin{table*}[t]
\caption{\label{tab:table}%
\setstretch{1.0}
Effect of social information on subjects' decisions.
The upper (lower) table summarizes the data for EXP-I (II). 
We divide the samples according to the size of $Z(i|r)$.
$N(\mbox{No.}|r)$ denotes the number of samples for case $r$ in each
 bin. 
$I(\mbox{No.})$ is the set of sample $i$ in each bin of case $O$
after removing the samples that satisfy $Z(i|O)<45\%$ or $Z(i|C)<(1-p(i))$ or
 $Z(i|M)<(1-p(i))$. 
 $|I(\mbox{No.})|$ means  
the number of samples in the set.
$p_{avg}$ is estimated as the average value of $p(i)=2(1-Z(i|O))$ 
over the samples in
$I(\mbox{No.})$. 
In the last two columns,
the ratio of the case with
$\{Z(i|r)<1/2\}$ for $r\in \{C,M\}$ among 
the samples in $I(\mbox{No.})$ is shown.
}
\begin{tabular}{rcccccccc}
No. & $Z(i|r)[\%]$ & $N(\mbox{No.}|O)$ & $N(\mbox{No.}|C)$ &
 $N(\mbox{No.}|M)$ & $|I(\mbox{No.})|$ & $p_{avg}(\mbox{No.})[\%]$ & $Z(i|C)<1/2$ & $Z(i|M)<1/2$ \\ 
\hline
1 &  $<5$ &       0    &  5  & 0  & NA& NA   & NA   & NA    \\
2 &  $5\sim 15$ & 3    & 33  & 7  & NA& NA   & NA   & NA    \\
3 &  $15\sim 25$& 5    & 28  & 25 & NA& NA   & NA   & NA    \\
4 &  $25\sim 35$& 18   &  9  & 30 & NA& NA   & NA   & NA    \\
5 &  $35\sim 45$& 35   &  5  & 13 & NA& NA   & NA   & NA    \\
6 &  $45\sim 55$& 38   &  5  & 13 & 38& 97.5 &18/38 & 17/38 \\
7 &  $55\sim 65$& 57   &  5  & 14 & 52& 78.3 & 7/52 &  5/52 \\
8 &  $65\sim 75$& 29   &  7  & 19 & 26& 60.3 & 0/26 &  0/26 \\
9 &  $75\sim 85$& 41   &  17 & 44 & 38& 40.6 & 0/38 &  0/38 \\
10 & $85\sim 95$& 11   &  57 & 62 & 11& 21.3 & 0/11 &  0/11 \\
11 & $\ge 95$    &  3  &  69 & 13 & 2& 5.1  &  0/2 &  0/2  \\
\hline 
Total & & 240 & 240 & 240 & 167 & 66.8\%  &  25/167 & 22/167 \\ 
\end{tabular}
\vspace*{0.3cm}
\begin{tabular}{rcccccccc}
No. & $Z(i|r)[\%]$ & $N(\mbox{No.}|O)$ & $N(\mbox{No.}|C)$ &
 $N(\mbox{No.}|M)$ & $|I(\mbox{No.})|$ & $p_{avg}(\mbox{No.})[\%]$ & $Z(i|C)<1/2$ & $Z(i|M)<1/2$ \\ 
\hline
1 &  $<5$ &       0   &  2    &  0  &NA& NA   & NA    & NA \\
2 &  $5\sim 15$ & 0   &  18   &  6  &NA& NA   & NA    & NA \\
3 &  $15\sim 25$& 8   &  22   & 18  &NA& NA   & NA    & NA \\
4 &  $25\sim 35$& 16  &  20   & 23  &NA& NA   & NA    & NA \\
5 &  $35\sim 45$& 36  &  8    & 16  &NA& NA   & NA    & NA \\
6 &  $45\sim 55$& 43 &  9    & 19  &43& 96.7 & 16/43 & 15/43 \\
7 &  $55\sim 65$& 46 &  10   & 16  &45& 79.3 & 8/45   & 3/45  \\
8 &  $65\sim 75$& 45 &  14   & 26  &45& 62.7 & 2/45   & 0/45 \\
9 &  $75\sim 85$& 33 &  33   & 56  &33& 41.9 & 0/33   & 0/33 \\
10 & $85\sim 95$& 11 &  67   & 54  &11& 21.3 & 0/11   & 0/11 \\
11 & $\ge 95$   &  2  &  37   & 6   &0& NA   & NA     & NA \\
\hline 
Total & & 240 & 240 & 240 & 177 & 68.7\% & 26/177 & 18/177 \\ 
\end{tabular}
\end{table*}

There are $240$ samples of sequences of choices for each $r$.
We divide these samples into 11 bins according to the size 
of $Z(i|r)$, as shown in Table \ref{tab:table}.
The number of data samples in each bin for
cases $r \in \{O,C,M\}$ are given in the second, third and fourth column
as $N(\mbox{No.}|r)$. 
Social information causes remarkable  changes in subjects' choices.
For case $O$, there  is one peak at No. 7, and for case $C (M)$, 
there are peaks at No. 2 (4) and No. 11 (10) in EXP-I. 
The samples in each bin of case $O$
share almost the same value of $p$.
For example,
 in the samples of No. 6 bin ($0.45<Z(i|O)\le 0.55$), there are almost only 
 herders in the subjects' sequence and $p(i) \simeq 100\%$. 
　In contrast, in the samples of No. 11 bin ($Z(i|O)>0.95$),
  almost all subjects know the answer to the questions and are
  independent $(p(i) \simeq 0\%)$. 
 An extremely small value of $Z(i|O)$ indicates some bias in the
 question and we omit the samples that satisfy $Z(i|O)<0.45$.
 In addition, the minimum value of $Z(i|r)$ should be  $1-p(i)$.
 If $Z(i|r)<1-p(i)$, it means that  
 the estimation of $p(i)$ for the sequence $\{X(i,t|r)\}$ fails.
 The true value of $p(i)$ should be larger than the estimated value. 
 We cannot give the appropriate estimation of $p(i)$ 
 for the choice sequence and we omit the samples that satisfy
 $Z(i|C)<1-p(i)$ or $Z(i|M)<1-p(i)$. 
 From these procedures, we are left with 167 (177) samples in EXP-I
 (II) and we denote the set by $I'$.
 $I(\mbox{No.})$ denotes the set of samples in 
 each bin in case $O$ among $I'$.
 
 We comment about the above data elimination procedure.
 The main purpose of the experiment is to clarify how herders copy 
 others' choices. For the purpose, it is necessary to assure that 
 herders choose each option with equal probability in case $O$ and 
 the herder's $q$ is 50\%. This is the precondition 
 of the experiment. We assume $q=0.5$ and  
 derive the above three conditions that $Z(i|r)$ should satisfy.
 If $Z(i|r)$ contradicts with at least one of the conditions,
 there is some bias in the options. The data for question $i$
 does not meet the precondition and we discard it in the analysis
 of the experimental data. The elimination procedure cannot assure
 the precondition with absolute certainty, it is indispensable.

We calculate the average value of $p(i)$ for the samples in 
$I(\mbox{No.})$.
We denote it as $p_{avg}(\mbox{No.})$ and estimate it as
\[
p_{avg}(\mbox{No.})=\frac{1}{|I(\mbox{No.})|}\sum_{i\in I(\mbox{\tiny{No.}})}p(i).
\]
Here,  $|I(\mbox{No.})|$  
in the denominator means the number of samples in $I(\mbox{No.})$, 
which is given  
in the sixth column of the table.
In the last two columns, we show the ratio of the case with
$\{Z(i|r)<1/2\}$ for $r\in \{C,M\}$ among 
the samples in $I(\mbox{No.}|O)$.
In both cases, as $p_{avg}$ increases, the ratio increases rapidly
to about half.

\begin{figure}[htbp]
\begin{center}
\includegraphics[width=7cm]{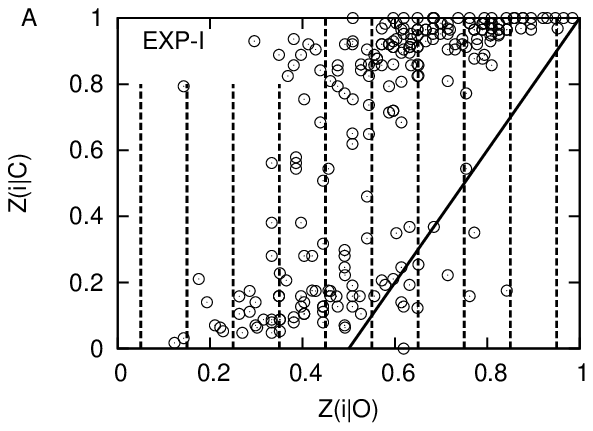}
\includegraphics[width=7cm]{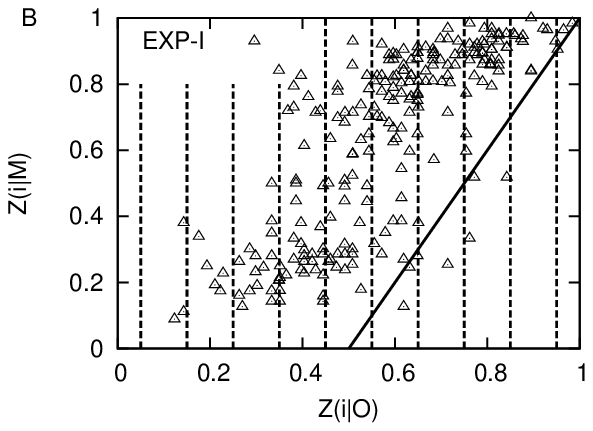}
\caption{\label{fig:scatter_Z}
\setstretch{1.0}
Scatter plots of $Z(i|O)$ vs. $Z(i|r)$ for (A) Case
 $C$ and (B) Case $M$. 
The vertical lines show the border of the bins
in Table \ref{tab:table}. 
The rising diagonal line from $(0.5,0)$ to top right shows the
 boundary condition $Z(i|r)=1-p$. 
}
\end{center}
\end{figure}			

 In order to see the social influence more pictorially, we 
 show the scatter plots of $Z(i|O)$ vs. $Z(i|r),r \in \{C,M\}$ of 
 EXP-I in Fig. \ref{fig:scatter_Z}. 
 The $x$-axis shows $Z(i|O)$ and  the
 y-axis shows $Z(i|r)$. The vertical lines 
 show the boundary between the bins (from No. 1 to No. 11) for case $O$
 in Table \ref{tab:table}.
 The rising diagonal line from $(0.5,0)$ to top right shows the
 boundary condition $Z(i|r)=1-p$. 
 If subjects' answers are not affected by  
 social information, data would distribute on the diagonal line
 from $(0,0)$ to top right. As
 the plots clearly indicate,
 the samples scatter more widely in the plane in case $C$ 
than in case $M$, which
 means that social influence is bigger in case $C$.
 For the samples with $Z(i|O)\ge 0.65$ in case $O$ (Nos. 8, 9, 10, and 11 bins in Table \ref{tab:table}), 
 the changes, $Z(i|C)-Z(i|O)$, are almost positive and $Z(i|C)$ takes a
 value of about 1 in case $C$.
 In case $M$, the changes, $Z(i|M)-Z(i|O)$, are also almost positive and $Z(i|M)$ takes a
 value of about 0.9.
 The average probability of choosing the correct option  
 improves with  social information for the
 samples in both cases. 
 In contrast, for 
 the samples with $0.45 \le Z(i|O)<0.65$ (Nos. 6 and 7 bins in Table \ref{tab:table}), 
 social information does not necessarily improve  
 average performance. There are many samples 
with $Z(i|r)-Z(i|O)<0$ in both cases.
 These samples constitute the lower  
 peak in Table \ref{tab:table}.

\subsection{Asymptotic behavior of the convergence}

 We have seen drastic changes in the distribution of $Z(i|r)$ 
 from the distribution of $Z(i|O)$.
 Table \ref{tab:table} and Figure \ref{fig:scatter_Z} show 
 the two-peak structure in the distribution of $Z(i|r)$. 
 In our previous work on the information cascade phase transition \cite{Mor:2012}, we
 have studied the time dependence of the
 convergence  behavior of the sequences $\{X(i,t|r)\}$.
\begin{figure}
\begin{center}
\begin{tabular}{c}
\includegraphics[width=7cm]{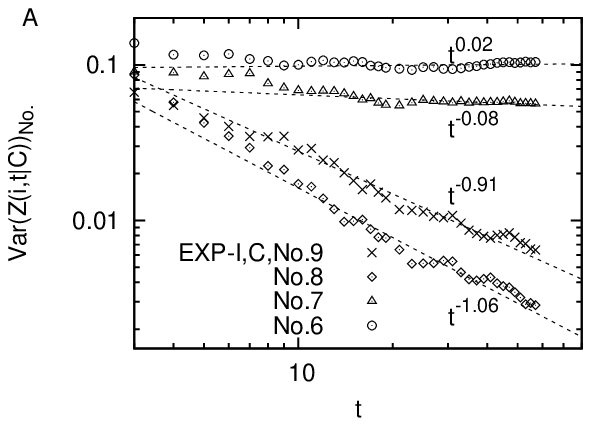} \\ 
\includegraphics[width=7cm]{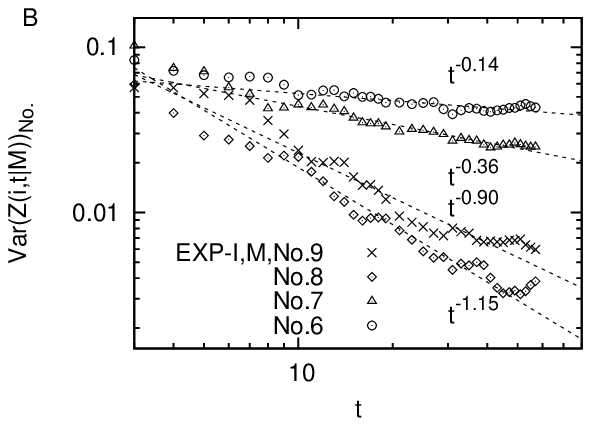}  \\
\includegraphics[width=7cm]{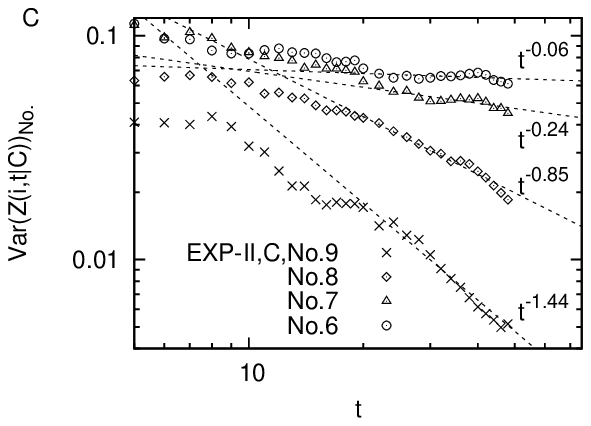} \\
\includegraphics[width=7cm]{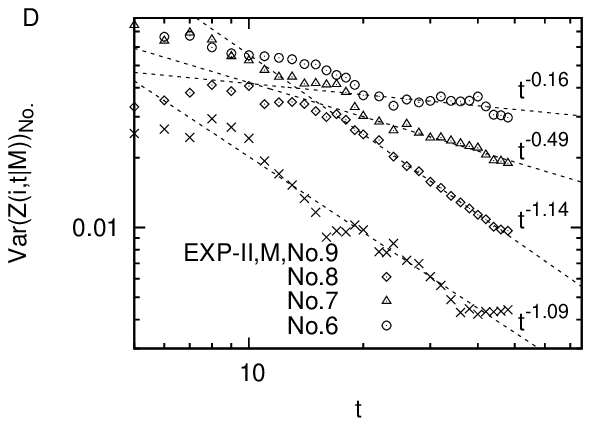}  \\
\end{tabular}
\caption{\label{fig:herd_macro}
\setstretch{1.0}
Convergent behavior.
Convergence is given by the  double logarithmic plot of 
$\mbox{Var}(Z(i,t|r))_{\mbox{No.}}$ vs. $t$ using the samples in four bins
 (Nos. 6 $(\circ)$, 7 $(\triangle)$, 8 $(\diamond)$, and 9 $(\times)$ in Table
 \ref{tab:table}) 
for  (A) Case $C$ in EXP-I, (B) Case $M$ in EXP-I, (C) Case $C$ in
 EXP-II, and (D) Case $M$ in EXP-II.
The dotted lines are fitted results with $\propto t^{-\gamma}$ for $t\ge
 10 (20)$ in EXP-I (II).}
\end{center}
\end{figure}

We denote the ratio of correct answers, $\frac{C_{1}(i,t|r)}{t}$, as 
\[
Z(i,t|r)\equiv \frac{C_{1}(i,t|r)}{t}=\frac{1}{t}\sum_{s=1}^{t}X(i,s|r).
\]
$Z(i,T_{i}|r)=Z(i|r)$ holds by definition. 
By studying the asymptotic 
behavior of the convergence of sequence $\{Z(i,t|r)\}$ for the samples in $I(\mbox{No.})$, 
one can  clarify
the possibility of  the information cascade transition by varying $p$.
The variance of $Z(i,t|r)$ for the samples in $I(\mbox{No.})$ is 
defined as 
\begin{eqnarray}
&&\mbox{Var}(Z(i,t|r))_{\mbox{No.}}\nonumber \\
&=&\frac{1}{|I(\mbox{No.})|}\sum_{i \in
I(\mbox{\tiny{No.}})}(Z(i,t|r)-<Z(i,t|r)>_{\mbox{No.}})^{2} \nonumber \\ 
&&<Z(i,t|r)>_{\mbox{No.}}=\frac{1}{|I(\mbox{No.})|}\sum_{i \in
I(\mbox{\tiny{No.}})}Z(i,t|r) \nonumber .  
\end{eqnarray}
Here, we denote the average value of $Z(i,t|r)$ over the samples in $I(\mbox{No.})$
by $<Z(i,t|r)>_{\mbox{No.}}$.
In the one-peak phase, the variance of $Z(i,t|r)$ for the
samples with the same  $p$  
converges to zero in thermodynamic limit $t\to \infty$. 
 In the analysis of experimental data, the
values of $p$ have some variance among the samples in each bin, and 
 Var($Z(i,t|r))_{\mbox{No.}}$ 
takes small values in the limit.  
 Depending on the convergence behavior, the one-peak phase is classified
 into two phases \cite{His:2012}.
 If Var($Z(i,t|r))_{\mbox{No.}}$ shows normal diffusive behavior as  
 $\mbox{Var}((Z(i,t|r))_{\mbox{No.}}\propto  t^{-1}$, it is called the 
normal diffusion phase. 
 We note that the variance is estimated for the ratio, $C_{1}(i,t|r)/t$, 
and the usual behavior $t^{1}$ for the sum of 
 $t$ random variables is replaced by $\propto t/t^{2}=t^{-1}$.
 If convergence is slow and $\mbox{Var}(Z(i,t|r))_{\mbox{No.}}\propto  t^{-\gamma}$
 with $0<\gamma<1$, it is called the super diffusion phase \cite{Hod:2004}.
 In the two-peak phase,  
 $\mbox{Var}(Z(i,t|r))_{\mbox{No.}}$ converges to some finite value in limit
 $t\to\infty$ \cite{His:2011}.

 Figure \ref{fig:herd_macro} shows the 
 double logarithmic plots of
 $\mbox{Var}(Z(i,t|r))_{\mbox{No.}}$  as a function of 
 $t$.
 We see that convergence becomes very slow as $p_{avg}(\mbox{No.})$ increases 
 in general.
 The convergence exponent $\gamma$ is estimated by fitting 
 with $\propto t^{-\gamma}$ for $t\ge 10$ in EXP-I.
 It decreases  almost monotonically  
 from  about 1 to $-0.02$ $(0.14)$ with an increase
 in $p_{avg}$ in case $C$ ($M$).   Taking into account the 
 estimate error of the exponent given in Appendix E,

 $\gamma$s are almost 1 for the samples in $I(9)$ and $I(8)$,
 and the system is in the normal diffusion (one-peak) phase in both
 cases $r \in \{C,M\}$. 
 For the samples in $I(7)$, $\gamma$s are apparently smaller than 1 and
 the system might be in the super diffusion phase.
 For the samples in $I(6)$, $\gamma$ becomes
 negative $(\gamma=-0.02)$ in case $C$. This suggests that 
 the system is in the two-peak phase for the samples in $I(6)$ \cite{Mor:2012}.
 In case $M$, $\gamma$ is positive even for the samples in $I(6)$
 and  the system
 might be in the super diffusion phase. However, the result does not
 necessarily deny the existence of the two-peak phase, taking into
 account the variance of $p(i)$ and the estimate error of $\gamma$
 from the limited sample size. 
 We can only say that if the two-peak phase exists, 
 the threshold value $p_{c}$ in case $M$
 is considerably larger than that in case $C$.

\section{\label{sec:analysis2}Data analysis: Microscopic Aspects}

In this section, we study the microscopic aspects of the herders.
We clarify how they copy others' choices and derive a microscopic 
rule in each case $r\in \{C,M\}$.  In particular, we study whether 
they behave as analog herders in case $M$.

\subsection{How do herders copy others?}

We determine how a herder's decision 
depends on social information. 
For this purpose, we need to subtract independent subjects'  
contribution from $X(i,t+1|r)$.
The probability of being independent is $1-p(i)$, and such a subject 
always chooses 1. A herder's contribution is estimated as
\[
( X(i,t+1|r)-(1-p(i))) /p(i).
\]
How the herder's decision depends on 
$C_{1}(i,t|r)=n_{1}$ is estimated by the expectation value of 
$(X(i,t+1|r)-(1-p(i))/p(i)$ under this condition.
The expectation value  means the probability that 
a herder chooses an option under the influence of 
prior $n_{1}$ subjects among $t$ who choose the same option.
We denote it by $q_{h}(t,n_{1}|r)$, and estimate it as 
\begin{equation}
q_{h}(t,n_{1}|r)
=\frac{\sum_{i \in
I'}\left[\frac{X(i,t+1|r)-(1-p(i))}{p(i)}\right]\delta_{C_{1}(i,t|r),n_{1}}}{\sum_{i\in
I'}
\delta_{C_{1}(i,t|r),n_{1}}}  \label{eq:q_h}.
\end{equation}
Here, $\delta_{i,j}$ is 1 (0) if $i=j \hspace*{0.2cm}(i\neq j)$ and the denominator 
 is  the number of sequences where $C_{1}(i,t|r)=n_{1}$. 
From the symmetry between $1\leftrightarrow 0$, we 
assume that $q_{h}(t,n_{1}|r)=1-q_{h}(t,t-n_{1}|r)$. 
We study the dependence of $q_{h}(t,n_{1}|r)$ on
$n_{1}/t$ and round $n_{1}/t$ to the nearest values 
in $\{k/13 (12) |k\in \{0,1,2,\cdots,13 (12)\}\}$ in EXP-I (II). 

\begin{figure}[h]
\begin{tabular}{c}
\includegraphics[width=7.5cm]{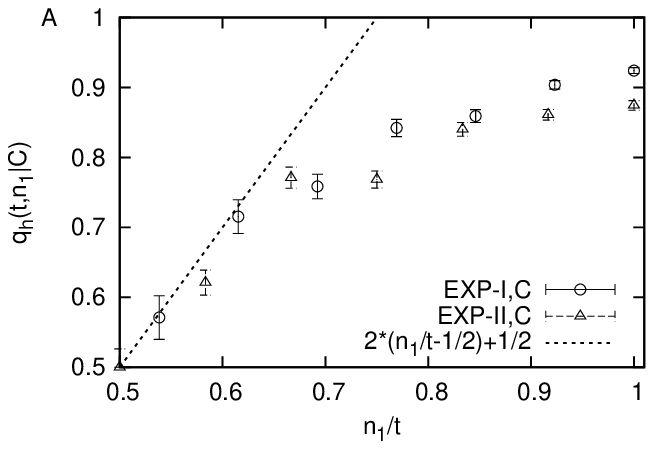} \\
\includegraphics[width=7.5cm]{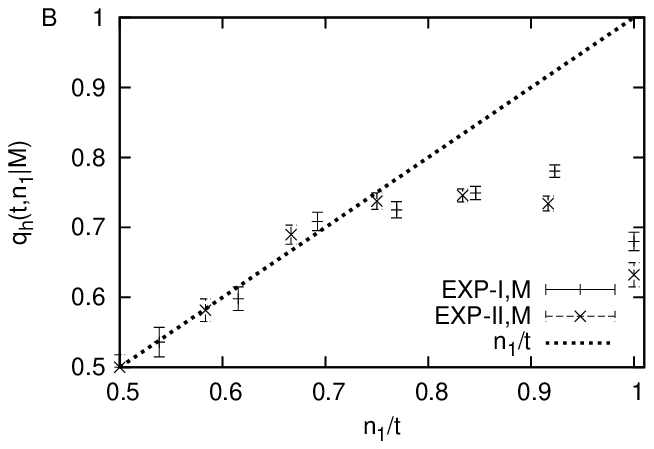}
\end{tabular}
\caption{\label{fig:herd_micro} 
\setstretch{1.0}
Microscopic rule of herder's decision for (A) Case $C$ and (B) Case $M$. 
It shows the probability $q_{h}(t,n_{1}|r)$ 
that a herder chooses an option under the influence of prior $n_{1}$ 
subjects among $t$ choosing that option in case $r$.
The thin dashed line in (A) shows $2(n_{1}/t-1/2)+1/2$.
 The dotted diagonal line in (B) shows the analog herder model 
$q_{h}(t,n_{1})=n_{1}/t$.
}
\end{figure}

Figure \ref{fig:herd_micro} 
shows the plot of $q_{h}(t,n_{1}|r)$ for (A) case $C$ and (B) case $M$.
We can clearly see the strong tendency to copy others in case $C$.
As $n_{1}/t$ increases from $1/2$, $q_{h}(t,n_{1}|C)$ rapidly increases and 
the slope at $n_{1}/t=1/2$ is about 2.0 in EXP-I.
 Such nonlinear behavior is known as a quorum response in social science
 and ethology \cite{Sum:2009}. 
 The magnitude of the slope measures the strength of the herders' response.
 Comparing EXP-I and EXP-II, the response of herders is more sharp in
 EXP-I than in EXP-II.  In EXP-II, where the amount of social  
information increases
 gradually, the subjects tend to copy others' choices
 more prudently than in EXP-I. 
 If the slope  exceeds 1, the system shows the information cascade
 transition. The transition ratio $p_{c}$  depends on the slope.
 In the digital herders case, where $q_{h}(t,n_{1})=\theta(n_{1}-t/2)$
 and the slope is infinite, $p_{c}$ takes $0.5$ \cite{His:2011}. 
 As the slope reduces to 1,
 $p_{c}$ increases to 1 and the phase transition disappears in the limit
 \cite{His:2010}.
 
 Contrary to case $C$, the dependence of $q_{h}(t,n_{1}|M)$ 
 on $n_{1}/t$ is weak and 
 the slope at $n_{1}/t=1/2$ is 
 almost 1 in case $M$. In range $1/4\le n_{1}/t \le 3/4$, 
  $q_{h}(t,n_{1}|M)$ lies on the diagonal dotted line and 
 the herders almost behave as analog herders.  
 As the multiplier $m$
 is the inverse of $n_{1}/t$ for a large $t$, the average herder  
 adopts the optimal max-min strategy in the range $4/3\le m \le 4$.
 As the slope at $n_{1}/t=1/2$ is small, if the information cascade phase
 transition occurs, the transition ratio 
 $p_{c}$ should become large as compared to in case $C$.  
 One can also see an interesting behavior of herders. If the 
 minority choice ratio $n_{1}/t $ is smaller than $1/4$ and 
 multiplier $m$ exceeds 4, some herders make the choice.
 As a result, if $n_{1}/t> 3/4$, $q_{h}(t,n_{1}|M)$ becomes almost
 constant, about $3/4$. We can interpret this as some of the herders preferring 
 a big multiplier (long-shot) and $q_{h}(t,n_{1})$ saturating at
 $3/4$.

\section{\label{sec:model}Analysis with  stochastic model}

In this section, 
we simulate the system 
by a stochastic model, which we call a voting model.
We consider a system with $p$ herders and $1-p$ independent voters.
We estimate the transition ratio $p_{c}$ and herder's
probability of correct choice in the experiment and  
compared it with that for the  analog herders system.

\subsection{Voting model and thermodynamic limit}

We introduce a stochastic process $\{X(t|p)\},t\in \{1,2,3,\cdots,T\}$
for $p \in [0,1]$.
$X(t+1|p) \in \{0,1\}$  is a Bernoulli random variable.
Its probabilistic rule depends on 
$C_{1}(t)=\sum_{t'=1}^{t}X(t'|r,p)$ and herders' proportion $p$.
Given $\{C_{1}(t)=n_{1}\}$, we denote the probability that 
a herder chooses (copies) the correct option by $q_{h}(t,n_{1})$.
As $q_{h}(t,n_{1})$ has symmetry $q_{h}(t,n_{1})=1-q_{h}(t,t-n_{1})$,
$q_{h}(t,n_{1})$ takes $1/2$ at $n_{1}/t=1/2$. 
We assume that $q_{h}(t,n_{1})$ is a smooth and monotonically increasing 
function of $n_{1}/t$.
The probabilistic rule  that
$X(t+1|r,p)$ obeys under the condition is
\begin{eqnarray}
\mbox{Prob}(X(t+1|p)=1|n_{1}) 
&=&(1-p)+p\cdot q_{h}(t,n_{1}), \nonumber \\
\mbox{Prob}(X(t+1|p)=0|n_{1})
&=&p\cdot (1-q_{h}(t,n_{1})). \nonumber  
\end{eqnarray}
We denote the  probability that $X(t+1|p)$ takes 1 under the
condition by $q(n_{1}/t|p)$ and 
the probability function Prob($C_{1}(t)=n)$ 
for $p$ by $P(t,n|p)$. 
The  master equation for $P(t,n|p)$ is
\begin{eqnarray}
P(t+1,n|p) 
&=&q((n-1)/t|p) \cdot P(t,n-1|p) \nonumber \\
&+& (1-q(n/t|p))\cdot P(t,n|p) \label{eq:master}. 
\end{eqnarray}
The expected value of $Z(t|p)=\frac{1}{t}C_{1}(t)$ is then estimated as
\[
\mbox{E}(Z(t|p))=\sum_{n=0}^{t}P(t,n|p)\cdot \frac{n}{t}.
\]

We are interested in the limit value of $Z(t|p)$ as $t\to \infty$, which 
we denote as $z$:
\[
z\equiv \lim_{t\to \infty}Z(t|p).
\]
In the one-peak phase, $Z(t|p)$ always converges to 
E($Z(t|p)$) in the limit, which we denote as $z_{+}$. 
In the two-peak phase, in addition to 
$z_{+}$, $Z(t|p)$ converges to a value smaller than half, which we denote 
 as $z_{-}$, with some positive probability. 
It is a probabilistic process and one cannot predict to which 
fixed point $Z(t|p)$ converges. 
To determine the 
threshold value $p_{c}$ between these phases and the limit value $z_{\pm}$,
 one needs to solve the following self-consistent equation \cite{His:2012}:
\begin{equation}
z=q(z|p)=(1-p)+ p \cdot q_{h}(t,t\cdot z) \label{eq:self}. 
\end{equation}
Given $p$, if there is only one solution, it is $z_{+}$ and  
the system is in the one-peak phase. The
 convergence exponent $\gamma$ is obtained by estimating
the slope of $q(z|p)$ at $z=z_{+}$ \cite{His:2012,Hod:2004}. 
If there are three solutions, which we denote as 
$z_{1}<z_{u}<z_{2}$, $z_{1} \hspace*{0.2cm}(z_{2})$ corresponds to
$z_{-} \hspace*{0.2cm}(z_{+})$. 
The middle solution $z_{u}$ is an unstable state and $Z(t|p)$ departs 
from $z_{u}$ as $t$ increases. 
The method gives the rigorous results for $z$ and $\gamma$
 where $q(z|p)$ is given as smooth function of $z$.

\begin{figure}[h]
\begin{center}
\includegraphics[width=6cm]{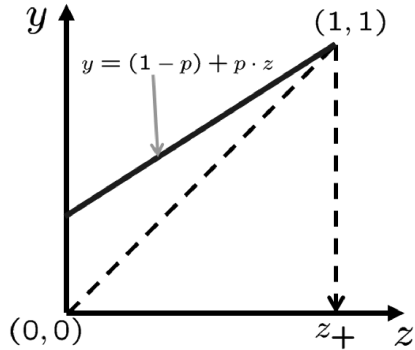} 
\caption{\label{fig:analog} 
\setstretch{1.0}
Schematic view of the self-consistent equation 
$z=q(z|p)=(1-p)+p\cdot q_{h}(t,t\cdot z)$
for the system of analog herders : $q_{h}(t,t\cdot z)=z$.
$(z,q(z|p))$ connects $(0,1-p)$ and $(1,1)$ by a direct line. 
There is only one stable solution $z_{+}$ at $z=1$ for $p<1$.
}
\end{center}
\end{figure}

Figure \ref{fig:analog} shows the case of analog herders and 
$q(z|p)=(1-p)+p\cdot q_{h}(t,t\cdot z)$ with $q_{h}(t,t\cdot z)
=z$ \cite{His:2010}. 
As one can easily see, for any value of $p<1$, 
there is only one stable solution $z_{+}$ at $z=1$. The system is in 
the one-peak phase and $Z(t|p)$ always converges to $z_{+}=1$ for $p<1$.
As the independent voter's probability of correct choice $q$ is 100\%,
that of herders is estimated as 1 by $(z_{+}-(1-p)\cdot 1)/p$.
Even in the worst limit $p\to 1$, the system of analog herders 
can  take the probability of correct choice to one.

\subsection{Transition ratio $p_{c}(r)$ for cases $r\in\{C,M\}$}

\begin{table}[htbp]
\caption{\label{tab:pc}
\setstretch{1.0}
Transition ratio $p_{c}$ of the voting (average herders)
 model. We determine $p_{c}$ using the condition that the self-consistent
 equation (\ref{eq:self}) has three or more solutions for $p>p_{c}$.}
\begin{tabular}{ccccc}
\hline
EXP. & $r$ &$p_{c}(r)$ & $r$ & $p_{c}(r)$ \\ 
\hline
I  & $C$ & 86.0\%  & $M$ & 95.7\% \\
II  & $C$ & 86.5\%  & $M$ & 96.7\% \\
\hline
\end{tabular}
\end{table}

We introduce an average herders model 
 where $q_{h}(t,t\cdot z)$ is given by linear extrapolation of the
 values $q_{h}(t,n_{1}|r)$ in equation (\ref{eq:q_h}).
In our previous work \cite{Mor:2012}, we model
 the behavior of herders  
by the following functional form with two parameters $a$ and $\lambda$:
\begin{equation}
\frac{1}{2} \left( a\tanh (\lambda(n_{1}/t-1/2))+1 \right). 
\label{q_model}
\end{equation}
However, the fitted result by the 
standard maximum likelihood estimation 
cannot capture the behavior of herders in the crucial 
region $n_{1}/t\sim 1/2$. 
We adopt the above linear extrapolated $q_{h}(t,n_{1}|r)$ for
$q_{h}(t,t\cdot z)$ and solve the 
self-consistent equation (\ref{eq:self}). We determine $p_{c}(r)$ for cases 
$r \in \{ C,M \}$ by the condition that the self-consistent equation has 
 three or more solutions. 
We summarize the results in Table \ref{tab:pc}.
In case $C$, $p_{c}(C)$ is 
from 86.0\% (EXP-I) to 86.5\% (EXP-II). 
In case $M$, $p_{c}(M)$ is from 95.7\% (EXP-I) to 96.7\% (EXP-II).
However, these estimates depend on the behavior of $q_{h}(t,n_{1})$ near 
$n_{1}/t=1/2$ where the estimate errors are big. We can at most say that 
$p_{c}(M)>p_{c}(C)$.   

\subsection{Herder's probability of correct choice}

\begin{figure}[htbp]
\begin{center}
\includegraphics[width=8cm]{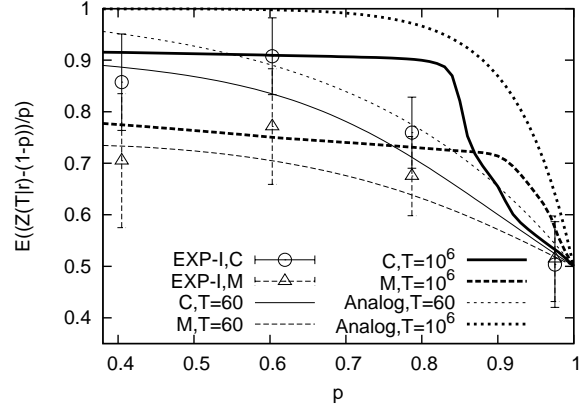}
\caption{\label{fig:p_vs_EHZ} 
\setstretch{1.0}
Plot of herders' probability of correct choice, $(\mbox{E}(Z(T|r))-(1-p))/p$ vs. $p$,
 for the voting model.
Symbol $\circ$ ($\triangle$) indicates the experimental data for 
the four bins $I(6),I(7),I(8)$, and $I(9)$ in Table \ref{tab:table} 
for case $C \hspace*{0.2cm}(M)$.
The lines show the 
results of the stochastic model with system size $T=60, r=C$ (thin solid); 
$T=60, r=M$ (thin dashed); $10^{6}, r=C$ (thick solid); and $10^{6},
 r=M$ (thick dashed).
We also plot the result of the stochastic model for analog herders
$q_{h}(t,n_{1})=n_{1}/t$ with $T=60$ (thin
 dotted) and $10^{6}$ (thick dotted).
}
\end{center}
\end{figure}

We estimate the probability of correct choice by a 
herder as a function of $p$ \cite{Cur:2006}. As for the voting model, it can be estimated 
using the expectation value of $Z(t|p)$ as
\[
\mbox{E}((Z(t|p)-(1-p)\cdot 1)/p). 
\]
For the experimental data, we 
take the average of $(Z(i|r)-(1-p(i))\cdot 1 )/p(i)$ over the samples in $I(\mbox{No.})$:
\[ 
\frac{1}{|I(\mbox{No.})|}\sum_{i \in I
(\mbox{\tiny{No.}})}(Z(i|r)-(1-p(i))\cdot 1)/p(i) .
\]
 We plot the results in Figure \ref{fig:p_vs_EHZ}.
 The experimental results show that the probability of correct choice in
 case $C$ is better than that in case $M$ except for the samples in $I(6)$. 
 As system size $T$ increases, for $p<p_{c}(C)$,
 the probability of correct choice in case $C$ remains better than that
 in case $M$.  
 However,
 the maximum value of $q_{h}(t,n_{1}|C)$ is about 0.9 and 
 the probability of correct choice saturates at the value for
 $p<p_{c}(C)$. 
 As $p$ exceeds $p_{c}(C)$, the probability of correct choice in case $C$ 
 rapidly decreases and dips below
 that in case $M$. From the information cascade transition, herders'
 probability of correct choice  
 is much lowered and this results in the poor performance.
 In contrast, the poor performance of herders in case $M$ for
 $p<p_{c}(C)$ comes from the saturation of $q_{h}(t,n_{1}|M)$ at
 $n_{1}/t=3/4$. 
 From the saturation, the
 probability of correct choice cannot reach
 the high value. For comparison, we show the results of the
 optimal system of analog herders with $T=60$ and $10^{6}$. In the thermodynamic
 limit, the probability of correct choice converges to one for $p<1$.

\section{\label{sec:conclusions}Conclusions}
 Social influence, which here is restricted only to information regarding 
 the choices of others, yields inaccuracy in the  majority
 choice.  
 If a herder receives summary statistics $\{C_{A},C_{B}\}$ and the
 payoff for the correct choice is constant, he strongly tends to 
 copy the majority. 
 The correct information given by independent voters 
 is buried below the herd and 
 the majority choice does not necessarily teach us the correct 
 one if herders' proportion exceeds $p_{c}(C)$ \cite{Mor:2012}.
 When the return is set to be 
 proportional to multipliers $\{M_{A},M_{B}\}$ that are
 inversely proportional to summary 
 statistics $\{C_{A},C_{B}\}$,
 the situation is a zero-sum game between a herder 
 and other previous subjects
 who have set the multipliers. 
 The optimal max-min strategy is that of analog herders 
 who choose  $\alpha \in \{A,B\}$ with probability proportional to
 $C_{\alpha}$.
 Furthermore, the system of analog herders  with $q=1$  
 maximizes the probability of the correct choice 
 for any value of $p$ in the thermodynamic limit.
 Even in limit $p\to 1$, only the system can take the
 probability of correct choice to one.

 We performed a laboratory experiment to
 study herders' behavior under the influence of 
 multipliers $\{M_{A},M_{B}\}$.
 We showed that they  collectively  behave almost
 as analog herders for $4/3 \le m \le 4 $, where $m$ is the multiplier.
 Outside the region, herders' copy 
 probability $q_{h}(t,n_{1}|M)$ 
 saturates at about $3/4$ for $n_{1}/t \ge 3/4$ and it deviates 
 from that of analog herders',
 $q_{h}(t,n_{1})=n_{1}/t$. As a result, the probability of 
 correct choice by a herder cannot reach a high value as compared 
 to in the system of analog herders. 

The system size and number of samples in our experiment are
very limited, and thus it is 
 difficult to estimate $p_{c}$ precisely.
More importantly, 
in the estimation of $p$, we assume herder's $q$ is 50\%.
It is the precondition of the experiment and we eliminate 
data which does not fulfill the condition.
However, the procedure does not 
assure the precondition.
In order to estimate $p$ more precisely and 
check the precondition, it is necessary to improve the 
experimental design or the data analysis procedure.
 In addition, in our experimental setup, the subjects have to
choose between A and B. In addition, in our analysis of the 
experimental data, we only observe the average behavior
of many herders. 
An interesting problem is whether a herder can adopt
the max-min strategy at the individual level
 or only the average herder can do it.
 In order to clarify this, one good way is to 
permit people to divide their choice 
 and vote fractionally. If the fraction voted by a 
subject is proportional to the summary
 statistic of previous subjects' choices, it suggests that the subject can 
 adopt the max-min strategy at the individual level. We think that a 
 more extensive experimental study
 of the system and of the related systems deserves further 
attention \cite{Sal:2006}.
 Such experimental studies should provide new approach to econophysics
 \cite{Man:2008,Lux:1995,Kir:1993,Con:2000,Gon:2011,Mor:2010} and 
 socio-physics\cite{Gal:2008}.

\begin{acknowledgment}
We thank Yosuke Irie  and Ruokang Han 
for their assistance in performing the experiment.
This work was supported by Grant-in-Aid for Challenging 
Exploratory Research 25610109.
\end{acknowledgment}

\bibliographystyle{jpsj}
\bibliography{65757}

\appendix

\section{\label{A}Additional information about the Experiment}

 In EXP-I, 120 subjects were recruited from the Literature 
 Department of Hokkaido University.
 We made two groups of about sixty subjects and 
the subjects in each group answered 120 questions 
one by one. Because of the capacity of the laboratory,  we 
could not perform the whole experiment at a time.
We divide the subjects of each group into five sub-groups of about 
12 subjects. In one session, subjects in a sub-group sequentially
 answered the questions. After five sessions
 we have gathered the data from all the subjects in a group.

 Subjects were paid in cash upon being released from the session.
 There was a 500 yen (about 5 dollars) participation fee and additional
 rewards that were proportional to the number
 of points gained.  In cases $O$ and $C$, one correct choice was 
 worth two points,
 and one point  was worth one yen (about one cents). 
 In case $M$, one correct choice was worth the
 multiplier itself.  In the main text, we treat case $M$ as zero-sum
 game. Considering the participation fee, we can regard it as constant 
 sum game, which  is equivalent to a zero-sum game. 
 As for EXP-II, detailed information can be 
 found in \cite{Mor:2012}.

\section{\label{PB}Experimental procedure}

  We explain the experimental procedure in EXP-I in detail.
  All the subjects in a sub-group 
  entered the laboratory and sat in the partitioned spaces.
  Using slides, we showed subjects how the experiment would proceed.
  We explained that we were studying how their choices were affected 
  by the choices of others.
  In particular, we emphasized  that 
  social information was realistic information calculated from the
  choices of previous subjects.
  Through the slides, we also explained how to calculate 
  multipliers $\{M_{A},M_{B}\}$
  in case $M$, with a concrete example.

  After the explanation, the subjects 
  logged into the experiment web site using their IDs and started
  to answer the questions.
  Interaction between subjects was permitted only through
  the social information given by the experiment server.
  A question was chosen by the experiment server and displayed on the
  monitor.  
  First, subjects answered the fist half of the 120 questions 
  $i \in \{1,2,\cdots,60\}$ 
  using only their own knowledge ($r=O$). 
  After answering all the sixty questions in case $O$, 
  the subjects answered the same 60 questions in case $C$.
  Finally, the subjects answered the same questions in case $M$.
  In each case, the experiment server chose a question among 
  the sixty questions
  at random that was not served to the other subjects at the time.
  Otherwise, we cannot give correct social information to the $t+1$-th
  subject from all previous $t$ subjects. 
  After a five-minute interval, we repeated the same procedure 
  so that the subjects answered all 120 questions.

\begin{figure*}[t]
\begin{center}
\includegraphics[width=13cm]{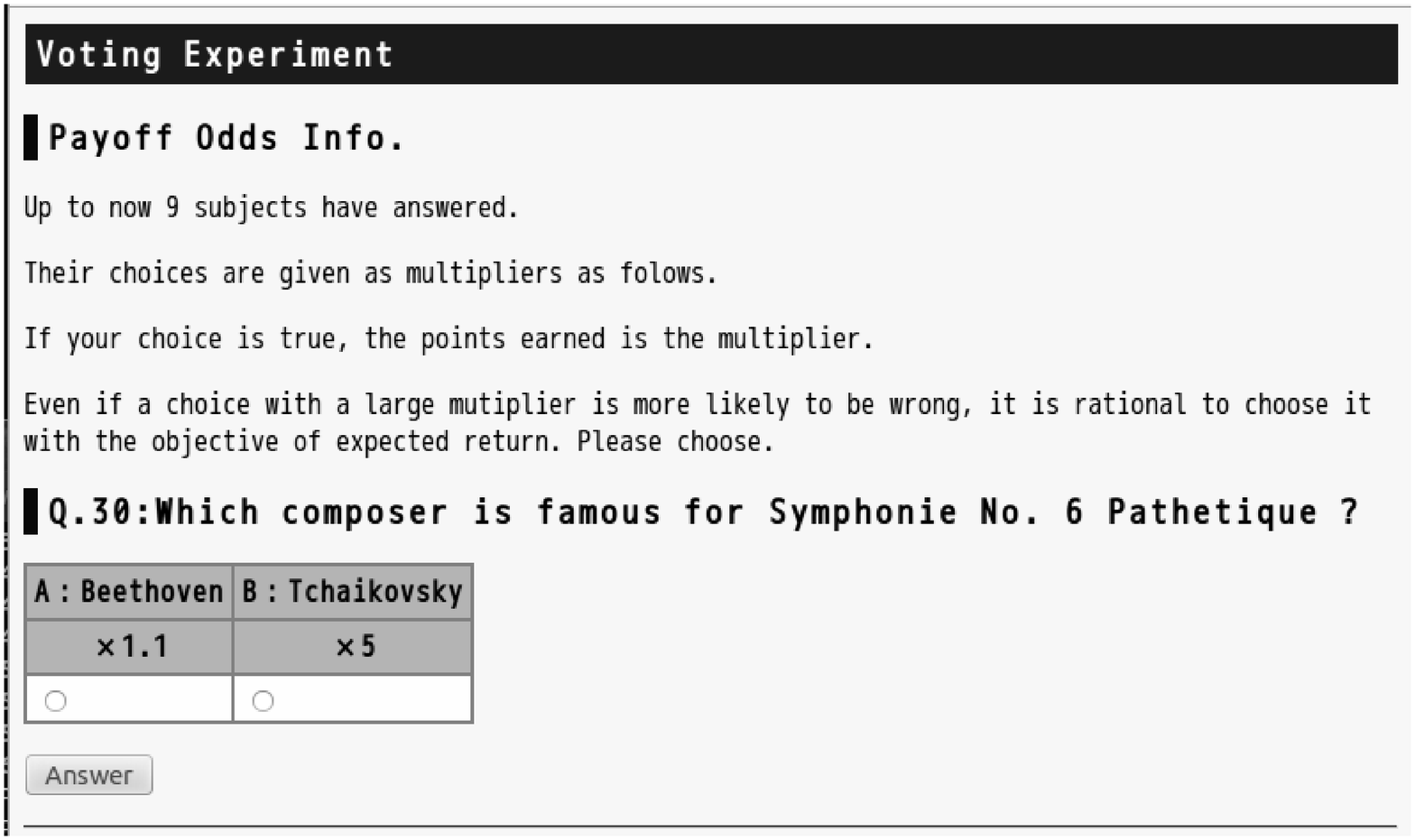} \\
\caption{\label{fig:experience}
\setstretch{1.0}
Snapshot of the screen for case $M$. 
Multipliers $\{M_{A},M_{B}\}$
are given in the 
second row in the box.}
\end{center}
\end{figure*}	

  Figure  \ref{fig:experience} shows the experience of the subjects in
  case $M$.
  In the example covered in the figure, already nine subjects 
  have answered question 30.
  The multipliers are given in the second row along with
  the number of subjects who answered the question.
  Only one subject among nine has chosen A and the remaining eight subjects
  have chosen B.
  Multiplier $M_{A}\hspace*{0.2cm}(M_{B})$ is calculated as 
  $10/(1+1)=5\hspace*{0.2cm}(10/(8+1)=1.1)$.
  The multipliers are rounded off to one decimal place.

  In EXP-II, the experience of the subjects is almost the same as
  in EXP-I \cite{Mor:2012}. 
  The difference lies in 
  how the experiment proceeds.
  In EXP-II, each subject answered each question from case $O$ to
  case $M$. After that, the experiment server chose another question.
  The process continues until the subject has answered all questions.
  The subjects were likely to easily remember the answers for the 
  earlier cases with different social information and 
  be careful in choosing 
  answers in the later cases. In order to exclude such an effect, we
  changed the system to that in EXP-I.

\section{\label{C}
Contorollability of the difficulty level of a question}
\begin{figure}[htbp]
\begin{center}
\includegraphics[width=7cm]{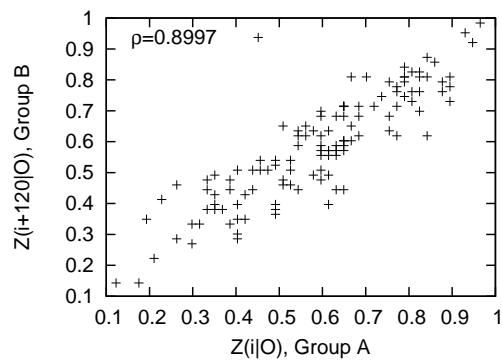}
\caption{\label{fig:compare}
\setstretch{1.0}
Scatter plots of $Z(i|O)$ vs. $Z(i+120|O)$ in EXP-I.
Pearson's correlation coefficient $\rho$ is $0.8997$.
}
\end{center}
\end{figure}

  We have used the same 120 questions in EXP-I and EXP-II. 
  For the selection process, please refer to our previous paper \cite{Mor:2012}.
  Here, we study whether the difficulty of a question is an inherent property 
  or not. For this purpose, we compare the percentage of 
  correct answers to each question in case O in Group A and in Group B.
  It is defined for Group A as $Z(i|O)=\sum_{s=1}^{T_{i}}X(i,s|O)/T_{i}$ and
  for Group B as $Z(i+120|O)=\sum_{s=1}^{T_{i+120}}X(i+120,s|O)/T_{i+120}$.
  We show the scatter plot 
  $\{Z(i|O),Z(i+120|O)\}$ in Figure \ref{fig:compare}.

 As one can clearly see the distribution almost on the diagonal line, 
 we can infer that there is a strong correlation. Pearson's correlation 
coefficient $\rho$ is about $0.90$. In EXP-II, we observe the same 
 feature and $\rho$ is about $0.82$.
 The strong correlation means that if a question 
 is difficult (easy) for the subjects in a group, it would also be 
 difficult (easy) for the subjects in the other group. 
The system sizes in our experiments are  very limited 
 and there remains some fluctuation in the estimation of $Z(i|O)$, 
 but it will disappear for a large system. 
 We can control the difficulty levels of the questions in the experiment and 
 study the response of a subject  under controlability.
 This aspect is important when one makes some prediction based on the 
 results presented in this paper.

\section{\label{D}Uniqueness of the analog herders  system}

In the main text, we show that the system of analog herders
maximizes the probability of correct choice for $p<1$ and 
can take it to one for any $p<1$.
Here, we show that 
only the system of analog herders can do it.

\begin{figure}[h]
\begin{center}
\includegraphics[width=7cm]{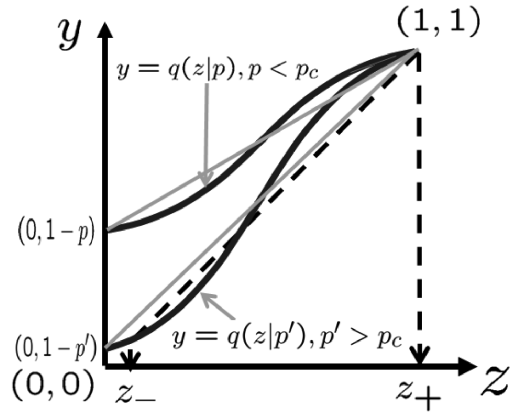}
\caption{\label{fig:general} 
\setstretch{1.0}
Schematic view of the self-consistent 
equation $z=q(z|p)=(1-p)+p\cdot q_{h}(t,t\cdot z)$
with general $q_{h}(t,t\cdot z)$.
$(z,q(z|p))$ connects $(0,1-p)$ and $(1,1)$ by a continuous 
curve. As $z_{+}=1$ is a stable solution, $\frac{q(z|p)}{dz}$
 at $z=1$
is one or less.
If $q_{h}(t,t\cdot z)$ deviates from $z$, for $p'>p_{c}$, 
in addition to the stable solution $z_{+}$ at $z=1$,
there is at least one stable solution $z_{-}$ for $z<1$.
}
\end{center}
\end{figure}

As the system with analog herders assures that the probability of correct
choice is one for any $p<1$, the self-consistent equation  for the 
system considered must have only one stable 
solution $z_{+}$ at $z=1$.
If the equation has more than one stable solution and the probability of
 convergence to solutions less than one is finite, the
 probability of correct choice 
 cannot take one. The self-consistent equation has a solution $z_{+}$ at $z=1$, 
$q_{h}(t,t\cdot z)$ must take 1 (0)  at $z=1 (0)$.
In addition, as $z_{+}$ is stable, the slope of $q(z|p)$
at $z=1$ is one or less. 
The curve $(z,q(z))$ connects 
$(0,(1-p))$ and $(1,1)$ as in Figure \ref{fig:general}. 
The curve of the system of analog herders 
connects the two points by a direct line.
If $q_{h}(t,t\cdot z)$ deviates from $z$, the curve
between the two points is rippling above and below the direct line.
Then, one can see that there is some threshold value $p_{c}<1$, where 
for $p>p_{c}$, the curve has more than three intersections with the diagonal 
line $y=z$. In this case, in addition to the stable solution $z_{+}$, there 
exists another stable solution $z_{-}$ less than one.
The probability of correct choice becomes less than one and 
the statement is proved.

\section{\label{E}Exponent $\gamma$}

\begin{figure}[htbp]
\begin{center}
\begin{tabular}{c}
\includegraphics[width=7cm]{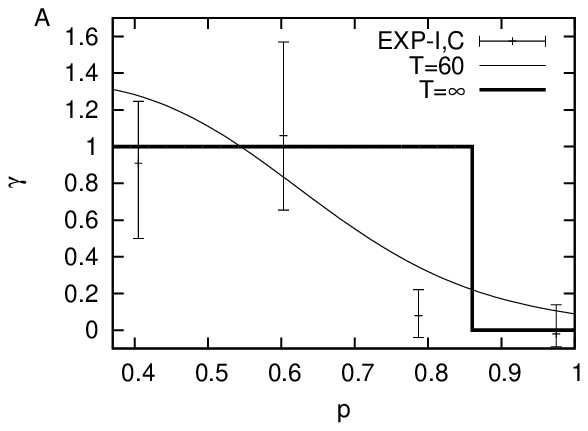} \\
\includegraphics[width=7cm]{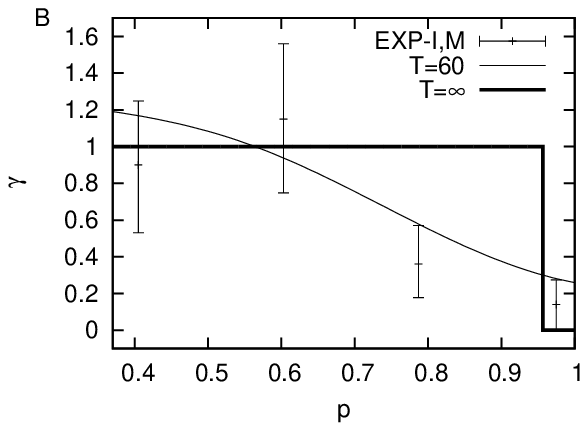}
\end{tabular}
\caption{\label{fig:p_vs_gamma}
\setstretch{1.0}
Plot of $\gamma$ vs. $p$.
We plot the results of the average herders model for EXP-I for (A) Case
 $C$ and (B) Case  $M$.
 Symbol ($\circ$) denotes $\gamma$s vs. $p_{avg}$ in EXP-I, which are 
estimated
in Figure \ref{fig:herd_macro}.
The lines show the 
results of the stochastic model with system size $T=60$ (thin
 solid) and  
$T=\infty$ (thick solid).}
\end{center}
\end{figure}		
In order to check the validity of the stochastic model
 for cases $r\in \{C,M\}$, we study the converge exponent $\gamma$. 
 We solve the master equation (\ref{eq:master}) recursively and obtain $P(t,n|p)$ for
$t\le T=60$ for EXP-I. 
We estimate the convergence exponent $\gamma$ from the slope
of $\mbox{Var}(Z(t|p))$ as
\begin{equation}
\gamma=\log \frac{\mbox{Var}(Z(T-\Delta T|p))}
{\mbox{Var}(Z(T|p))}/\log \frac{T}{T-\Delta T}.
\end{equation}
We take $\Delta T=50$ to match
the analysis of the experimental data in Figures
\ref{fig:herd_macro}A and B.
In order to give  
the error bar of $\gamma$ for the experimental results, 
 we adopted the voting model to simulate the system and 
 estimate  the 95\% confidence interval \cite{Mor:2012}.
For $T=\infty$ (thermodynamic limit), 
we estimate the gradient $q'(z_{+}|p)$ of $q(z|p)$ at $z=z_{+}$ 
and use the formula $\gamma=\mbox{Min}(1,2-2\cdot q'(z_{+}|p))$
\cite{His:2012}.
The results are summarized in Figure \ref{fig:p_vs_gamma} for (A) case $C$
 and (B) case $M$. 
For $T=60$, the model describes 
the experimental results well.
In the limit $T\to \infty$, $\gamma$ 
monotonically decreases from 1 to 0.

\end{document}